# The association between stratospheric weak polar vortex events and cold air outbreaks

Erik W. Kolstad & Tarjei Breiteig

Bjerknes Centre for Climate Research, Bergen, Norway

Adam A. Scaife

Hadley Centre for Climate Prediction and Research, Met Office, Exeter, United Kingdom

**Abstract**

13 Previous studies have identified an association between near-surface temperature anomalies in
14 the Northern Hemisphere and weak stratospheric polar westerlies. Large regions in northern
15 Asia, Europe and North America have been found to cool in the mature and late stages of
16 stratospheric weak vortex events. A substantial part of the temperature changes are associated
17 with changes to the tropospheric Northern Annular Mode and North Atlantic Oscillation pressure
18 patterns. The apparent coupling between the stratosphere and the troposphere may be of
19 relevance for weather forecasting, but only if the temporal and spatial nature of the coupling is
20 known. Here we show, using 51 winters of re-analysis data, that the tropospheric temperature
21 development relative to stratospheric weak polar vortex events goes through a series of well-
22 defined stages, including geographically distinct cold air outbreaks. At the inception of weak
23 vortex events, a precursor signal in the form of a strong high-pressure anomaly is found over
24 Northwest Europe. At the same time, long-lived and robust cold anomalies appear over Asia and
25 Western Europe. A few weeks later, near the mature stage of weak vortex events, a shorter-lived
26 cold anomaly emerges off the east coast of North America. The probability of cold air outbreaks
27 in different phases of the weak vortex life cycle increases by 50–80 % in four key regions. This
28 shows that the stratospheric polar vortex contains information that can be used to enhance
29 forecasts of cold air outbreaks. 300-year pre-industrial control runs of 11 state-of-the-art coupled
30 climate models corroborate our results.

31

31

## 1 Introduction

33 Cold air outbreaks (CAOs) are departures of cold air masses into warmer regions. Over land,
34 these events can lead to excess deaths and damage to infrastructure. Over the ocean, marine
35 CAOs are important for a number of reasons: they give rise to mesoscale weather phenomena
36 such as polar lows (Bracegirdle and Gray, 2008), they lead to enhanced heat and momentum
37 fluxes from the ocean to the air (Renfrew and Moore, 1999) and may therefore influence the
38 ocean circulation (Pickart et al., 2003), and they cause rapid formation of sea ice in marginal ice
39 zones (Skogseth et al., 2004).

40 In recent years it has emerged that anomalies in the stratospheric circulation can be associated
41 with tropospheric CAOs (Thompson et al., 2002; Cai and Ren, 2007; Scaife et al., 2008).
42 Normally, the extra-tropical stratosphere is characterized by a strong westerly circumpolar flow.
43 In winter, planetary waves of tropospheric origin continuously propagate into the stratosphere
44 (Charney and Drazin, 1961), where they break and exert a drag on the zonal flow (McIntyre and
45 Palmer, 1983). This violates geostrophic balance and induces a poleward drift of air masses. At
46 high latitudes, the air converges, sinks and warms adiabatically. If there is more wave-breaking
47 than normal, the stratospheric zonal flow weakens and the polar stratosphere warms, giving rise
48 to Stratospheric Sudden Warmings (SSWs; Matsuno, 1971). These conditions may last for days
49 to weeks (Limpasuvan and Hartmann, 1999).

50 After their first appearance in the upper stratosphere, circulation anomalies are occasionally
51 found at successively lower levels (Matsuno, 1970; Lorenz and Hartmann, 2003). After reaching
52 the tropopause, the anomalies may impact the troposphere through an interaction with synoptic-scale eddies, or more directly through induced meridional circulations (Song and Robinson, 2004). The result are negative Northern Annular Mode (NAM; Thompson and Wallace, 2001) and North Atlantic Oscillation (NAO; Hurrell et al., 2003) patterns near the surface some weeks after the first warming signal in the upper stratosphere.

Negative NAM and NAO regimes in the troposphere have a profound influence on the weather in large and widespread regions of the Northern Hemisphere. Atlantic and Pacific storm tracks shift latitudinally (Hurrell and Van Loon, 1997; Baldwin and Dunkerton, 2001), Greenland and Newfoundland warms (Thompson et al., 2002), and the frequency and severity of CAOs increase over large parts of East Asia (Jeong and Ho, 2005), northern Eurasia and the north-western part of North America (Thompson and Wallace, 2001; Walsh et al., 2001; Cellitti et al., 2006). Over the ocean, negative phases of the NAO, and positive height anomalies over Greenland in particular, have been found to be associated with marine CAOs over the Nordic Seas (Kolstad et al., 2008).

Motivated by the observed link between anomalous stratospheric events and the tropospheric climate, we aim to provide a detailed description of tropospheric cold anomalies relative to such events. Thompson et al. (2002) investigated the mean temperature response in the first 60 days after the onset dates of stratospheric anomalous vortex conditions. Here we extend their work by assessing the temperature development and changes to the probability of CAOs at different stages of stratospheric weak vortex events. We do so for both continental and oceanic regions.

We find that the tropospheric temperature development goes through several distinct and well-defined stages of stratospheric weak vortex events. CAOs over both continental and oceanic regions are identified. These results are corroborated by data from 300-year time slices of 11 coupled model runs.

## 2 Data and methods

Daily mean fields from the NCEP/NCAR re-analysis (NNR) data (Kalnay et al., 1996) were used throughout the study. The analysis period was from the autumn/winter of 1958 to the winter/spring of 2009.

Monthly mean data from 11 models in the World Climate Research Programme's Coupled Model Intercomparison Project phase 3 (CMIP3) multi-model dataset were also used. The models' originating groups and countries, abbreviations, horizontal and vertical resolutions, highest pressure levels and the model years that were used in this study are listed in Table I. For each model, 300-year time slices of the «pre-industrial control run», with no anthropogenic or natural forcing, were used. These were chosen arbitrarily from the years that were available for download.

The most commonly used measure of stratospheric variability is the NAM index. However, as both the spatial structure and the temporal variability of the NAM differed greatly across the models, we used a Vortex Strength Index (VSI), defined below, as our indicator of stratospheric variability. 50 hPa was the highest pressure level for which data from all the models was available, so this was used as the stratospheric reference level for both the models and NNR.

The daily VSI was computed from NNR as follows. For each winter (December–February; DJF) each day is defined by its date $D$ and its year $Y$. $\Phi_{D,Y}$ is defined as the daily area-averaged geopotential height at 50 hPa north of 65°N. From this time series, the date-wise climatological mean $\mu_D^\Phi$ and standard deviation $\sigma_D^\Phi$ were computed for 31-day windows surrounding the dates, 97   so that $\mu_D^\Phi = \sum_y \sum_{d=D-15}^{D+15} \Phi_{d,y}$ and $\sigma_D^\Phi = \sqrt{\sum_y \sum_{d=D-15}^{D+15} (\Phi_{d,y} - \mu_D^\Phi)^2}$. The daily, dimensionless VSI is given

98   by $-(\Phi_{D,Y} - \mu_D^\Phi)/\sigma_D^\Phi$. The minus sign is there because strong vortices are characterized by

99   negative height anomalies. The monthly VSI was computed from the model data in a similar

100  way, without the smoothing.

101  The analysis in this paper is centred on composites of days and months for which the

102  stratospheric vortex is weak. A *weak vortex day* in NNR is a day for which the daily VSI falls

103  below its overall $10^{th}$ percentile. Similarly, a *weak vortex month* in the models is a month for

104  which the monthly VSI is less than its overall $10^{th}$ percentile.

105  Weak vortex day number $i$ in NNR is referred to as $d_0^i$, where $i$ is an integer between 1 and $N$,

106  the total number of such days. With 51 winters, each consisting of 90 days, $N$ is 459. We define

107  an N-*composite* as an average over all the $N$ cases. The $N$-composites of daily tropospheric

108  anomalies were computed as follows. If $Z'(d_j^i)$ is the $i$th geopotential height anomaly $j$ days

109  *after* (if $j$ is positive, or *before* if $j$ is negative) the weak vortex days, the $N$-composite height

110  anomaly in the time interval from 0 to 20 days after the weak vortex days is

111  $\overline{Z}'\big|_0^{20} \equiv \frac{1}{21 \times N} \sum_{i=1}^{N} \sum_{j=0}^{20} Z'(d_j^i)$. The same procedure is used for $N$-composite temperature anomalies

112  $\overline{T}'$.

113  The statistical significance of the $N$-composite anomalies was computed by means of Monte

114  Carlo experiments. Each $d_j^i$ is defined at a specific date $D$ and in a specific year $Y$. For

115  geopotential height, 500 artificial $\hat{Z}'_k$, for which the *years* of the $d_j^i$ were permuted at random

116  from the analysis period, while the original *dates* were kept fixed, were constructed. With this

117  method, the seasonal cycle of the original *N*-composites is retained. Furthermore, to preserve the

118  autocorrelation between the $d_j^i$, any $d_j^i$ from the same year in the original *N*-composite were

119  given the same year in the $\hat{Z}'_k$. An *N*-composite anomaly $\bar{Z}'$ is considered significantly different

120  from climatology (at the 0.05 level) if it was lower than the 0.025 quantile or greater than the

121  0.975 quantile of the $\hat{Z}'_k$.

122  Four domains that illustrate the spatial extent of tropospheric cold anomalies are defined in Table

123  II. For each of these domains, $T_{D,Y}$, area-averaged daily time series of 700-hPa temperature were

124  computed. These time series were adjusted for seasonality and standardized in the same way as

125  the VSI above, i.e. the temperature index is given by $(T_{D,Y} - \mu_D^T)/\sigma_D^T$. A *CAO day* is a day for

126  which this index is lower than its overall 10$^{th}$ percentile.

127

## 3    The Daily Vortex Strength Index

In Figure 1(a), a matrix of VSI values for each day in the analysis period is shown. The blue days are the $d_0^i$. Although the figure shows that weak vortex events can last for weeks, all the $d_0^i$ are considered as separate incidents in this study. An alternative would be to decide on a key date for each event, e.g. the day that the VSI starts declining or the day that it hits its minimum. That methodology was used by Charlton and Polvani (2007), and their central dates of SSWs are shown with crosses in Figure 1(a). An advantage of such an approach is that all the events are independent, and the study of lead/lag processes is free of effects of artificial smoothing. However, a disadvantage is that one must be certain that the correct reference date has been chosen in each case. Our approach is sensitive to only one a priori choice: the selection of a threshold value for weak vortex days.

As our method leads to an artificial smoothing of the temporal signal, we average over rather long time intervals and adopt the terminology of Limpasuvan et al. (2004). They examined the evolution of wave activity fluxes and atmospheric pressure fields in several sub-periods of SSW life cycles, and defined the following phases of weak vortex events: *onset* (60–41 days before the $d_0^i$), *growth* (40–21 days before the $d_0^i$), *early mature* (20–1 days before the $d_0^i$), *late mature* (0–20 days after the $d_0^i$), *decline* (21–40 days after the $d_0^i$) and *decay* (41–60 days after the $d_0^i$). Note that the developments in Figure 9 in Limpasuvan et al. (2004) are not necessarily directly comparable to the developments in our time intervals.

147    In Figure 1(b), the *N*-composite average of the VSI index is shown with up to 60 days of

148    lag/lead-time relative to the $d_0^i$. The slopes of the curve on either side of day zero are indicators

149    of the autocorrelation of the VSI near weak vortex events.

150

## 4 Results

### 4.1 Composite analysis of daily data

In this section, *N*-composites of daily 500-hPa geopotential height and temperature (from NNR) anomalies relative to the $d_0^i$ are analyzed. The temporal development in the troposphere throughout the life cycle of stratospheric weak vortex events is presented with an emphasis on cold anomalies. Figure 2 shows the development of 500-hPa geopotential height anomalies averaged over the weak vortex stages specified in Section 3. Figure 2 provides the dynamical framework to interpret the 700-hPa potential temperature anomalies in Figure 3.

In the *onset* phase (Figure 2(a)), a positive anomaly centred over northern Scandinavia (H0 in Figure 2) and a negative anomaly near the Bering Strait are found. This corresponds to a pattern found to favour stratospheric warmings through upward-propagating tropospheric waves (Kuroda and Kodera, 1999; Breiteig, 2009; Garfinkel et al., 2009) and is a tropospheric precursor of the warming aloft.

H0 is present throughout all the stages and moves westwards with time. In Figure 2(a), two additional negative height anomalies are found, one over Northeast Asia (L1) and another over the western Mediterranean (L2). These two anomalies constitute a wave together with H0. L1 and L2 persist for all the stages without migrating substantially. A third negative height anomaly (L3) emerges to the south of Newfoundland in Figure 2(b).

In Figure 3, three distinct and persistent cold anomalies are found: C1 over Northeast Asia is associated with L1; C2 at different locations over Europe and the Nordic Seas is associated with L2; and C3 off the coast of North America is associated with L3.

172  In addition, there is an early cold anomaly in the western part of the Bering Sea and the Sea of
173  Okhotsk in the *growth* phase (Figure 3(b)) that is associated with the Bering Sea negative height
174  anomaly.

175

176  4.2    Model ensemble analysis

177  We now examine monthly mean data from the 11 coupled climate models in Table I. Figure 4
178  shows the 700-hPa temperatures during weak vortex months (b), as well as in the preceding (a)
179  and succeeding (c) months. For comparison, parts of the temperature development from the re-
180  analysis (Figure 3) are shown again, but now with averaging periods corresponding to the ones
181  for the models.

182  The following features are present in both the model results for the months before weak vortex
183  months (Figure 4(a)) and in NNR 16–45 days prior to the $d_0^i$ (Figure 4(d)): a cold anomaly over
184  the Bering Sea and the Sea of Okhotsk, C1 over Asia, C2 over Europe, and warm anomalies over
185  the Nordic Seas region and northern North America.

186  The westward shift of the Nordic Seas warm anomaly, the northward spread of C2, and the
187  appearance of C3 off the coast of North America in the month surrounding the $d_0^i$ (Figure 4(e))
188  are all seen in the model ensemble during weak vortex months (Figure 4(b)). The pattern formed
189  by these anomalies, along with the warm anomaly near the Caspian Sea, is reminiscent of the
190  typical NAO quadrupole temperature pattern (Stephenson and Pavan, 2003). In the same time
191  period, C1 is more zonally elongated in the models than in NNR, covering parts of the Northwest
192  Pacific, and C3 covers more of North America.

193 In the late stages of weak vortex event life cycles, C2, now centred over Scandinavia in NNR
194 (Figure 4(f)), is present in the models (Figure 4(c)). C1 over Asia is weaker in the models than in
195 NNR at this stage.

196 In summary, the most pronounced cold anomalies in Figure 3 are well matched by cold
197 anomalies in an ensemble of 11 coupled climate models.

198

199 4.3     Regional cold anomalies

200 In this section, we use time series of area-averaged and standardized 700-hPa temperature
201 anomalies to assess whether or not information about the state of the stratospheric polar vortex
202 has an impact on the forecasting of CAOs in the four regions in Table II.

203 CAO days were defined above as the days with temperatures lower than the overall 10th
204 percentile. We now define the ratio $r$ as the total number of CAO days in each stage of a weak
205 vortex event divided by the climatological mean number of CAO days on the same dates of the
206 year throughout the analysis period. When $r > 1$ for a specific stage, CAOs are more likely than
207 normal during that stage of weak vortex events.

208 In Figure 5, $N$-composites of (non-standardized) temperature anomalies relative to the $d_0^i$ (a) and
209 the ratio $r$ (b) are shown for each region and for the different stages of weak vortex events.

210 In the NEA domain (green curves in Figure 5), the East Asia winter monsoon ridge-trough
211 pattern (Compo et al., 1999) in Figure 2 maintains a steady flow of cold air into the region and
212 lower-than-normal temperatures are found throughout the life cycle of weak vortex events
213 (Figure 5(a)). In the *growth*, *early mature* and *decay* stages, CAOs occur about 70–80 % more often than normal (Figure 5(b)), with temperatures between 1 K and 1.4 K lower than normal. The *decline* phase sees about 50 % more CAOs and temperatures 0.9 K lower than normal.

In the EUR domain (red curves in Figure 5), CAOs are about 50 % (40 %) more likely than normal in the *growth* (*decline*) phase, with temperatures about 0.7 K lower than normal. This is consistent with C2 covering large parts of this region in Figure 3(b) and 3(e).

As C2 moves northward into the NS domain in the *late mature* phase (Figure 3(d)), L2 expands towards the north (Figure 2(d)), and H0 weakens the climatological trough west of Iceland. An anomalous ridge-trough pattern emerges between Greenland and Central Europe, and the flow field veers towards the north. The interplay between an anomalous ridge over Greenland/Iceland and an anomalous trough over Scandinavia was found by Kolstad et al. (2008) to be associated with marine CAOs over the Nordic Seas. Indeed, CAOs are about 60 % more likely than normal and temperatures are about 1 K lower than normal in the *late mature* and *decline* phases (black curves in Figure 5).

In WNA (blue curves in Figure 5), the lowest temperatures are found in the *mature* phase. Around this time the development of L3 and the ridging in the Baffin Island region (Figure 2(c)–(d)) are symptoms of a southward displacement of the 500-hPa «polar vortex» in Cellitti et al. (2006). This situation is associated with substantial cold air advection and C3 in Figures 3(c)–(d). The chance of CAOs in the region increase by about 40 % in the *early mature* phase and about 60 % in the *late mature* phase of stratospheric weak vortex events. Temperatures are 0.6 K and 1 K lower than normal, respectively.

## 5 Concluding remarks

The relationship between stratospheric weak vortex events and tropospheric developments, and cold air outbreaks (CAOs) in particular, were investigated by using 51 winters of re-analysis data. The life cycle of a weak vortex event was separated into six 20-day periods (following Limpasuvan et al., 2004): the *onset*, *growth*, *early mature*, *late mature*, *decline* and *decay* phases.

The clearest precursor of stratospheric weak vortex events was found to be a high pressure anomaly centred over northern Scandinavia in the *onset* and *growth* phases. This positive height anomaly persisted for all the phases and was contained to the high latitudes in the Atlantic sector.

A major motivation for studying the troposphere-stratosphere interactions is the prospect of prediction. A ratio $r$ was defined as the total number of CAO days in each stage of a weak vortex event divided by the climatological mean number of CAO days on the corresponding dates. The probability of CAOs was found to increase:

- by about 70–80 % in the *growth*, *early mature* and *decay* phases and by about 50 % in the *decline* phase in Northeast Asia.

- by about 50 % in the *growth* phase and by about 40 % in the *decline* phase in Central Europe.

- by about 60 % in the *late mature* and *decline* phases in northern Scandinavia, including the Nordic Seas.

- by about 40 % in the *early mature* phase and by about 60 % in the *late mature* phase off the east coast of North America.

255 Parts of the analysis were repeated with an ensemble of 11 coupled climate models. Somewhat
256 surprisingly, considering that many of these models do not have a well-resolved stratosphere, the
257 model results corroborated the relationships between weak vortex events and the cold anomalies
258 listed above. This may indicate that the main aspects of the tropospheric temperature
259 development during the life cycle of stratospheric weak vortex events are associated with
260 internal processes in the troposphere and lower stratosphere.

261

**Acknowledgements**

Erik Kolstad's work was funded by the Norwegian Research Council through its International Polar Year programme and the project IPY-THORPEX (grant number 175992/S30). Adam Scaife was supported by the Joint DECC, Defra and MoD Integrated Climate Programme - DECC/Defra (GA01101), MoD (CBC/2B/0417_Annex C5). This is publication no. X from the Bjerknes Centre for Climate Research.

| Originating group(s) | Country | IPCC model designation | Horizontal resolution north of 20N | Number of vertical layers | Number of layers over 200 hPa | Top level | St. dev. of DJF 50-hPa geop. height (m) | Model years used |
|---|---|---|---|---|---|---|---|---|
| Bjerknes Centre for Climate Research | Norway | BCCR-BCM2.0 | 25x128 | 31 | 20 | 0.01 hPa | 102 | 2370–2669 |
| National Center for Atmospheric Research | USA | CCSM3 | 50x256 | 26 | 13 | 2.2 hPa | 208 | 480–779 |
| Canadian Centre for Climate Modelling & Analysis | Canada | CGCM3.1(T63) | 25x128 | 31 | N/A | N/A | 179 | 1850–2149 |
| Météo-France / Centre National de Recherches Météorologiques | France | CNRM-CM3 | 25x128 | 45 | 23 | 0.05 hPa | 148 | 2130–2429 |
| Max Planck Institute for Meteorology | Germany | ECHAM5/MPI-OM | 37x192 | 31 | 9 | 10 hPa | 118 | 2250–2549 |
| US Dept. of Commerce / NOAA / Geophysical Fluid Dynamics Laboratory | USA | GFDL-CM2.1 | 35x144 | 24 | 5 | 3 hPa | 153 | 199–498 |
| Institut Pierre Simon Laplace | France | IPSL-CM4 | 28x96 | 19 | 8 | N/A | 184 | 2060–2359 |
| Center for Climate System Research / National Institute for Environmental Studies / Frontier Research Center for Global Change | Japan | MIROC3.2(medres) | 25x128 | 20 | 8 | 30 km | 96 | 2500–2799 |
| Meteorological Research Institute | Japan | MRI-CGCM2.3.2 | 25x128 | 30 | 16 | 0.4 hPa | 190 | 1901–2200 |
| National Center for Atmospheric Research | USA | PCM | 25x128 | 26 | 13 | 2.2 hPa | 170 | 150–449 |
| Hadley Centre for Climate Prediction and Research / Met Office | United Kingdom | UKMO-HadCM3 | 29x96 | 19 | N/A | N/A | 151 | 1900–2199 |

Table I. The coupled climate models used in this study. More information and relevant references for each model can be found at http://www-pcmdi.llnl.gov/.

| Region name | Abbreviated region name | Longitude range | Latitude range |
|---|---|---|---|
| Nordic Seas | NS | 10°W–60°E | 62.5°N–80°N |
| Central Europe | EUR | 10°W–30°E | 40°N–57.5°N |
| Western North Atlantic | WNA | 90°W–40°W | 30°N–47.5°N |
| Northeast Asia | NEA | 80°E–140°E | 42.5°N–65°N |

Table II. The domains used in this study.

**Figure Captions**

Figure 1. (a) The standardized daily VSI, sorted and binned by quantiles. The SSW central dates in Charlton and Polvani (2007) are marked with crosses. (b) The *N*-composite standardized daily VSI (in quantile terms) for each day relative to weak vortex days.

Figure 2. *N*-composites of 500-hPa geopotential height anomalies (in m) relative to weak vortex days, averaged over the specified phases. Anomalies that are significant at the 0.05-level, according to a 500-member Monte Carlo experiment (see text for details), are marked with dots. Black contours are drawn along the values corresponding to the tick marks of the colour bar.

Figure 3. As Figure 2, but for 700-hPa temperature anomalies (in K). The boundaries of the four regions in Table II are drawn with black boxes.

Figure 4. (a–c): The model ensemble 700-hPa temperature anomalies (in K) relative to weak vortex months. (d–f) As Figure 3, but in the specified time intervals relative to weak vortex days.

Figure 5. For the four regions in Table II, the *N*-composite (a) area-averaged, non-standardized 700-hPa temperature anomalies relative to weak vortex days; and (b) number of CAO days divided by the climatological number of CAO days (the ratio *r*).

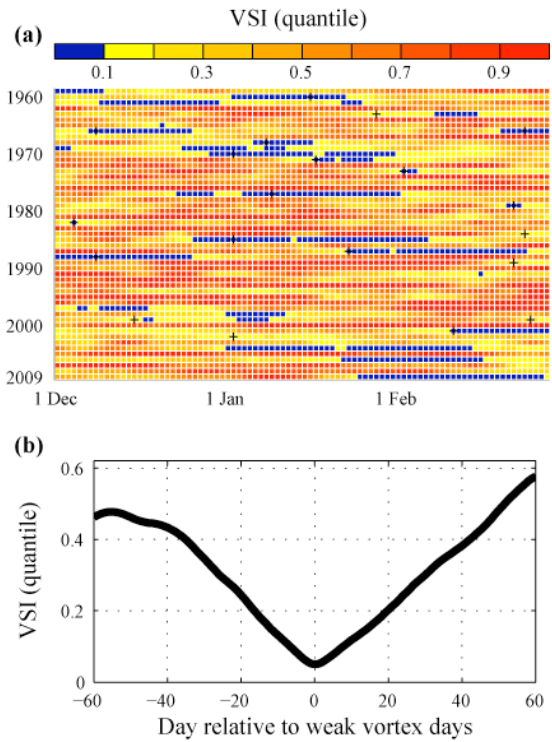

Figure 1. (a) The standardized daily VSI, sorted and binned by quantiles. The SSW central dates in Charlton and Polvani (2007) are marked with crosses. (b) The *N*-composite standardized daily VSI (in quantile terms) for each day relative to weak vortex days.

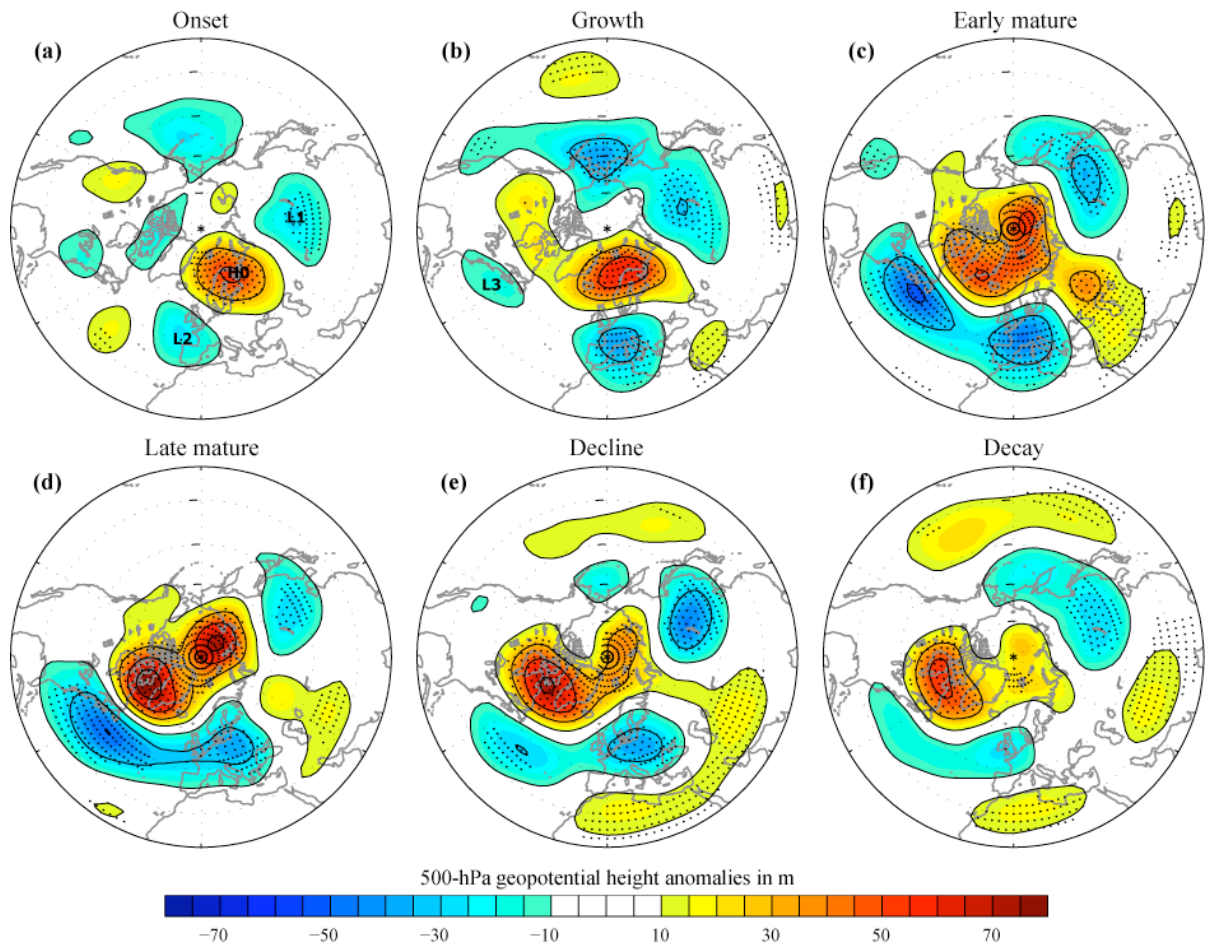

Figure 2. *N*-composites of 500-hPa geopotential height anomalies (in m) relative to weak vortex days, averaged over the specified phases. Anomalies that are significant at the 0.05-level, according to a 500-member Monte Carlo experiment (see text for details), are marked with dots. Black contours are drawn along the values corresponding to the tick marks of the colour bar.

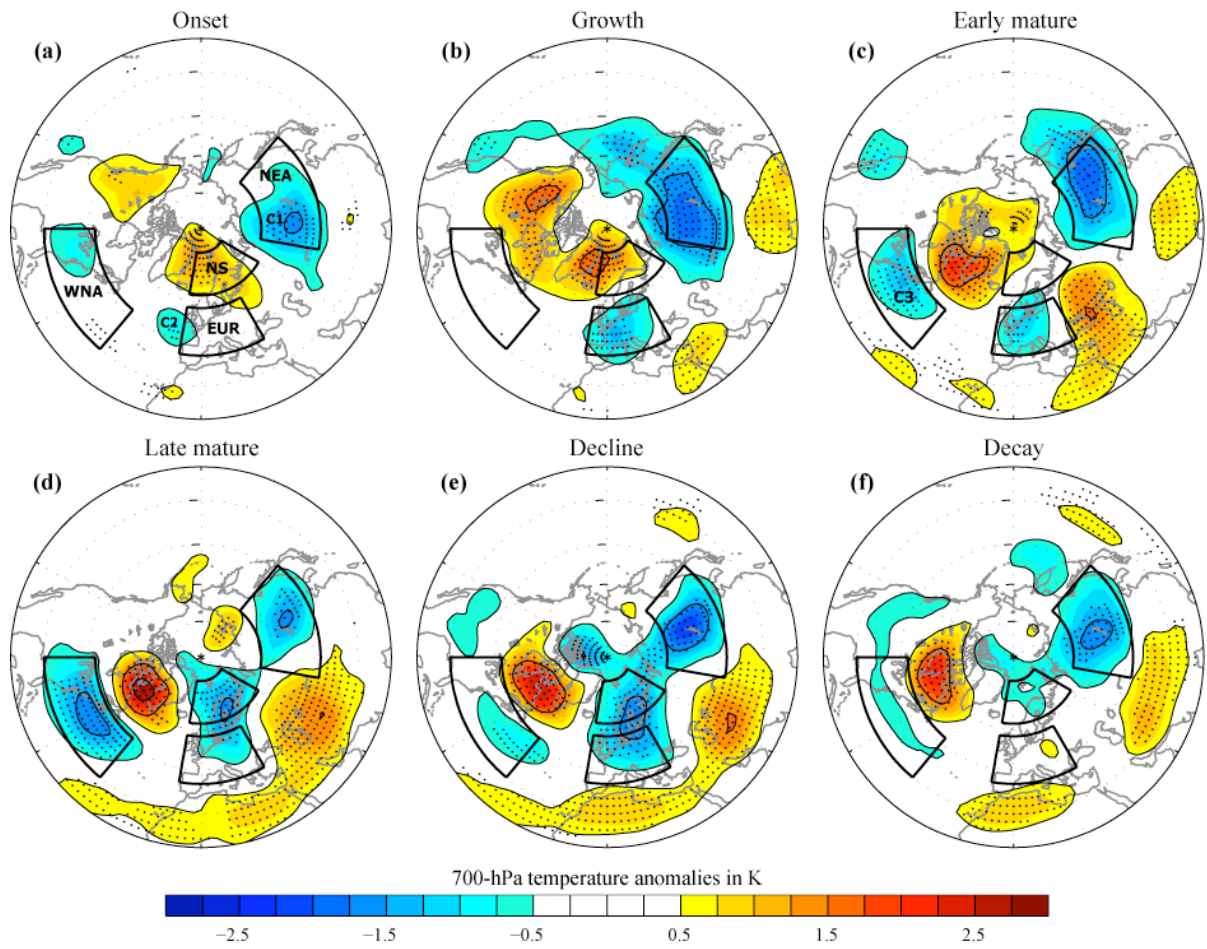

Figure 3. As Figure 2, but for 700-hPa temperature anomalies (in K). The boundaries of the four regions in Table II are drawn with black boxes.

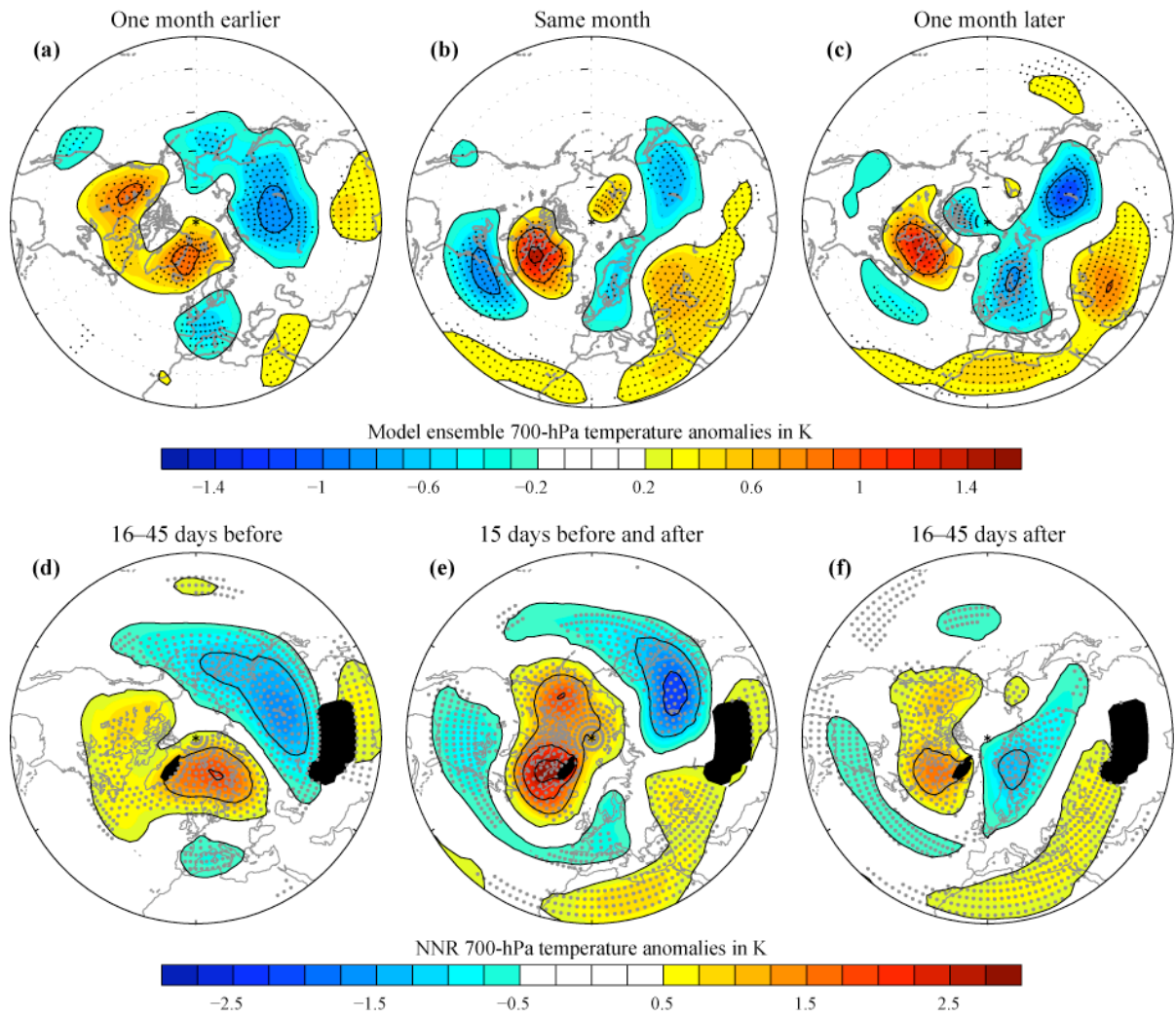

Figure 4. (a–c): The model ensemble 700-hPa temperature anomalies (in K) relative to weak vortex months. (d–f) As Figure 3, but in the specified time intervals relative to weak vortex days.

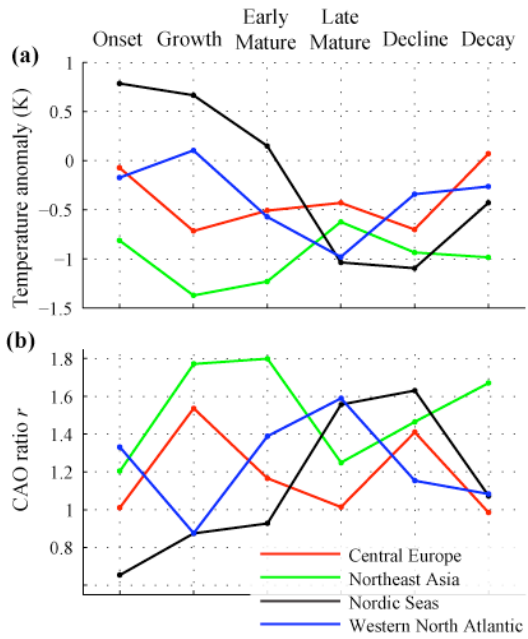

Figure 5. For the four regions in Table II, the *N*-composite (a) area-averaged, non-standardized 700-hPa temperature anomalies relative to weak vortex days; and (b) number of CAO days divided by the climatological number of CAO days (the ratio *r*).